\documentclass[aip,reprint,amsmath,amssymb]{revtex4-1}

\usepackage{graphicx}
\usepackage{dcolumn}
\usepackage{bm}

\usepackage[utf8]{inputenc}
\usepackage[T1]{fontenc}
\usepackage{enumitem}
\usepackage{tgschola}
\usepackage{hyperref,physics,caption,subcaption}
\usepackage{etoolbox}
\usepackage{comment}

\makeatletter
\def\@email#1#2{%
 \endgroup
 \patchcmd{\titleblock@produce}
  {\frontmatter@RRAPformat}
  {\frontmatter@RRAPformat{\produce@RRAP{*#1\href{mailto:#2}{#2}}}\frontmatter@RRAPformat}
  {}{}
}%
\makeatother

\begin{document}
\preprint{AIP/123-QED}

\title{
Thermodynamics of Driven Systems via the Kuramoto-Sivashinsky Equation}

\author{E. Hansen}%
\affiliation{Institute of Fusion Studies, University of Texas at Austin}
 \email{ehansen99@utexas.edu}
 
\author{W. Barham}
\affiliation{Theoretical Division, Los Alamos National Laboratory}
 
\author{P.J. Morrison}
\affiliation{Institute of Fusion Studies, University of Texas at Austin}
 \homepage{https://web2.ph.utexas.edu/~morrison/}

\date{\today}

\begin{abstract}
We examine the differences between the driven turbulence described by the Kuramoto-Sivashinsky (KS) equation and the second law of thermodynamics.
A general velocity and entropy density system is analyzed with the unified thermodynamic algorithm of metriplectic dynamics, and we show that the positive spectra of the KS equation due to an external energy source prevent its metriplectic description.
A variant of the KS equation is produced that monotonically generates an entropy, but the only equilibria of this variant system are spatially constant.
Numerical experiments are performed comparing the evolution of the KS equation and its thermodynamic variant.
The entropy of this thermodynamic system is increased further by the driving effects of the KS equation, reconciling the generation of entropy with the energy source of the KS equation.
Further numerical experiments restrict the positive spectra in the KS equation to determine the effect on the system time evolution. 
While rescaling the growth rates of instabilities reproduces similar behavior on a slower time scale, introduction of individual positive spectra reproduces the formation of equilibria, relative equilibria, and a transition to chaos.
\end{abstract}

\maketitle

\section{\label{sec:intro}
Introduction}

\noindent The second law of thermodynamics is commonly associated with the universal increase of disorder.
In a fluid context, this is interpreted as the uniform distribution of fluid properties. \cite{KunduCohen}
But many examples of physical systems exist where patterns emerge spontaneously and nontrivial solutions form. 
We mention four such examples here.
In the study of thermal combustion, flame fronts exhibit complicated motion.
Wrinkles form through the interaction of thermal expansion and diffusion, and their amplitude is limited by turbulence in the gas.\cite{Sivashinsky1977}
In the study of plasmas in fusion reactors, an instability known as the trapped ion mode develops spontaneously and has major implications for plasma confinement. 
Steady states of the trapped ion mode appear as oscillations which are excited by ion-electron collisions and depleted by the process of Landau damping. \cite{LMRT}
In the study of thin films flowing down an inclined plane, wavy patterns form due to a competition of gravitational acceleration and viscous dissipation.\cite{Benney66,Homsy1974,Nepomnyashchii1974}
In the study of chemical turbulence and reaction diffusion systems, certain reaction equilibria are given energy by diffusion, and higher order viscous damping allows formation of a stable pattern.\cite{KuramotoTsuzuki1976}
In each of these examples, we observe a competition between the dissipation expected from the second law of thermodynamics and mechanisms which cause oscillations to grow.
We seek an understanding of how the entropy develops in a setting where we see the formation of patterns.
\vspace{\baselineskip}\newline
\noindent Model systems describing each of the above examples have been analyzed asymptotically by separating long and short wavelength behavior.
In all cases, this leads to the same reduced model, the Kuramoto-Sivashinksy (KS) equation, \cite{HNZOrder}
\begin{align}
        \frac{\partial v}{\partial t} + vv_x = - v_{xx} - \nu v_{xxxx} \,, \hspace{1cm} 0\leq x\leq L \,.
\end{align}
Here $v(x,t)$ is a dependent variable depending on the position $x$ in the spatial domain and time $t$.
Throughout this study, we consider the spatial domain to be periodic.
In the case of combustion, $v(x,t)$ defines the gradient of a perturbation to a flame front surface. 
In the case of the trapped ion mode, $v$ represents a scaled electric potential.
In the case of thin films, $v$ represents the stream function.
In the case of chemical turbulence, $v$ represents the gradient of the amplitude phase.
\vspace{\baselineskip}\newline
\noindent The KS equation is able to account for instabilities through the second derivative term and dissipation through the hyperviscosity term, $v_{xxxx}$, and the balance of these terms coupled with the nonlinear convective derivative defines the behavior of the system. 
Moreover, it is simple enough to apply the tools of nonlinear dynamics which are often applied to discrete maps.
As the length scale of the KS equation increases, nontrivial standing waves develop, which undergo bifurcations into traveling waves and higher wavenumber cellular solutions.\cite{GreeneKim,FrischCells,MichelsonStates,HooperGrimshaw}
One also observes the formation of chaos and strange attractors as wave solutions develop, and the fact that many Fourier modes are dissipated has allowed estimates of the dimension of this strange attractor.\cite{NicolaenkoGlobal,HNZOrder,KassamTrefethen,CDE10}
Thus, the KS equation exhibits a wide variety of behavior required of systems with both forcing and dissipation, allowing the equation to serve as a useful model of driven turbulence.
\vspace{\baselineskip}\newline
\noindent Here, we examine how the nonlinear dynamics and pattern formation modeled by the KS equation are consistent with the growth of universal entropy.
This may be accomplished with the metriplectic formalism, which describes systems that satisfy the first and second laws of thermodynamics.\cite{Morrison1984,Morrison1986,MorrisonUpdike}
Specifically, metriplectic systems use a symmetric bracket (which implies a degenerate gradient flow) to add dissipative terms to an energy-conserving, Hamiltonian description.
The dissipation monotonically generates some quantity conserved by the Hamiltonian dynamics, thus turning a thermodynamically reversible system into an irreversible one.
The recent unified thermodynamic (UT) algorithm uses the principles of nonequilibrium thermodynamics to provide choices of bivectors for a particular system that construct this symmetric bracket. 
Several systems have been equipped with a metriplectic bracket in this way, including the Navier-Stokes-Fourier system of a fluid with heat conduction.\cite{BarhamMorrisonZaidni24,ZaidniMorrison}
When the dissipative and forcing higher derivatives are removed, the KS equation matches Burgers' and the 1D Navier-Stokes equations, and hence satisfies the first law of thermodynamics in the same way.
With this similarity to a metriplectic fluid system, we question whether we can use the UT algorithm to find the generated entropy we need for the KS equation.
\vspace{\baselineskip}\newline
\noindent As we will show, the answer is negative precisely because of the driving that causes the presence of instability.
However, due to the small scale dissipation of the KS system, it is possible to build an entropy for a metriplectic KS equation which only includes dissipation.
We then perform a series of numerical experiments with this metriplectic KS system.
The body of literature around the KS equation is vast, and provides many examples of steady and traveling solutions, the formation of chaos from wave interactions, and energy generation and depletion.
These examples are tested in the metriplectic KS system to determine what if any chaotic behavior is retained.
We may also use the evolution of the KS equation to determine what would happen to the positive-definite metriplectic entropy if unstable modes are driven.
The modifications removing unstable modes also inspire a new approach to varying the instabilities of the KS equation, and we test what solutions form when all positive spectra vary in amplitude or only certain modes are retained.

\section{\label{sec:metriplectic}
Metriplectic Structure of KS Equation}

\noindent As mentioned earlier, the UT algorithm relates the dissipative dynamics of a metriplectic system to the thermodynamic fluxes $J^\alpha$ and forces $Z_\alpha$ of that system.
Consider a system with velocity $v$ and entropy density $\sigma$.
The UT algorithm then prescribes the dynamics of the system as \cite{ZaidniMorrison}
\begin{align}
    \frac{\partial v}{\partial t} - \{v,H\} &= \mathcal{L}^v(J^v) \,, \\ \nonumber\frac{\partial \sigma}{\partial t}-\{\sigma,H\} &= \mathcal{L}^\sigma(J^\sigma)+Z_\alpha \tilde{L}^{\alpha\beta}Z_\beta \,,  
\end{align}
where $\{,\}$ is a Poisson bracket, 
\begin{align}
    H = \int_0^L \bigg(\frac12 v^2 + \sigma T \bigg)dx
\end{align}is the total energy of the system,\cite{Morrison98} $\tilde{L}^{\alpha\beta}$ is a tensor of phenomenological coefficients defining the system dissipative dynamics, and $\mathcal{L}^{\alpha}$ is a pseudodifferential operator.
Below the symbol $H_{\alpha} = {\delta H}/{\delta \alpha}$ is the functional derivative of $H$ with respect to the field $\alpha$.
If the forces and fluxes are taken as 
\begin{align}
    Z_{\alpha} = \mathcal{L}^\alpha_* H_{\alpha} \quad \textrm{and} \quad  J^\alpha = -H_{\sigma} \tilde{L}^{\alpha\beta}Z_\beta \,,
\end{align}
with $\mathcal{L}_*^\alpha$ being the dual of the operator $\mathcal{L}^\alpha$, the total energy of the system will be conserved.\cite{ZaidniMorrison}
The UT algorithm defines a metriplectic bracket of the system as the Kulkarni-Nomizu product \cite{MorrisonUpdike}
\begin{align}
    \int_0^L &\Sigma(dF,dG)M(dK,dN)+\Sigma(dK,dN)M(dF,dG) \\ -&\Sigma(dF,dN)M(dG,dK)-\Sigma(dG,dK)M(dF,dN) \nonumber dx
\end{align}
of the bivectors 
\begin{align}
     \Sigma(dF,dG) &= (\mathcal{L}_*^\alpha F_\alpha  )\tilde{L}^{\alpha\beta}(\mathcal{L}_*^\beta G_\beta) \,,
     \\ \nonumber
     M(dF,dG) &= F_{\sigma}G_\sigma
     \, .
\end{align}
In what follows we will attempt to define an entropy such that the KS equation coupled with that entropy is a metriplectic system.
To do this, we want to recover entropy advection, so we modify the Burger's equation Lie-Poisson bracket to \cite{Morrison98}
\begin{align}
    \{F,G\} = \int_0^L &-\frac13 v (F_v \partial_x G_v - G_v \partial_x F_v) \\ \nonumber &- \sigma (F_v \partial_x G_\sigma - G_v \partial_x F_\sigma ) dx \,.
\end{align}
Notice that the total entropy $S = \int_0^L \sigma dx$ satisfies $\{F,S\} = 0$ for an arbitrary functional $F$, as $S$ is required by the metriplectic formalism to be a Casimir invariant. \cite{MorrisonUpdike} 
Aside from the entropy advection, we don't want to modify any dynamics of the KS equation, so we take the phenomenological coefficients to be zero with the exception of $\tilde{L}^{vv}$.
Then the UT algorithm gives a system
\begin{align}
    \frac{\partial v}{\partial t} + vv_x &= -\mathcal{L}^v(T \tilde{L}^{vv}\mathcal{L}_*^v(v)) \,,\label{eq:genks} \\ \nonumber
    \frac{\partial \sigma}{\partial t} + \partial_x(\sigma v) &= \tilde{L}^{vv} (\mathcal{L}_*^v(v))^2 \,,
\end{align}
so long as the temperature is taken to be constant. (This allows a one-way coupling from the velocity of the KS equation into an entropy).
\vspace{\baselineskip}\newline
\noindent We observe that the operator on the right hand side of the velocity equation in \eqref{eq:genks} has a negative spectrum under the assumptions $T>0 $, $\tilde{L}^{vv} > 0$ since $\mathcal{L}^v\mathcal{L}^v_*$ is positive for an arbitrary choice of the pseudodifferential operator.
However, the operator $T_{KS} \equiv -\partial_{xx} -\nu \partial_{xxxx}$ has a positive element of its spectrum, since upon acting on an element of Fourier space $e^{ikx}$, the result is $(k^2 - \nu k^4)e^{ikx}$, and $k^2 - \nu k^4 \geq 0$ for sufficiently small $k \leq k_c \equiv \sqrt{\nu^{-1}}$ .
This demonstrates that the instabilities of the KS equation for long wavelengths prevent the definition of an entropy through the UT algorithm.
\vspace{\baselineskip}\newline
\noindent In order to maintain a thermodynamically consistent component of the KS equation, we see that we will have to remove the unstable modes of the system.
If we write a velocity profile 
\begin{align} \nonumber
    v(x) = \sum_{k=-\infty}^\infty  v_k  e^{2\pi i kx /L}
\end{align}
in terms of its Fourier components $v_k$, the result of applying the operator $T_{KS}$ to $v(x)$ is given by
\begin{align}
    T_{KS}v(x) = \sum_{k=-\infty}^\infty  (k^2 - \nu k^4)v_k  e^{2\pi i kx /L} \,.
\end{align}
To exclude instabilities from the KS system, we restrict the effect of $T_{KS}$ to modes which are depleted, defining a filtered dissipation operator
\begin{align}
    \hspace{-0.5em} T_{diss}v(x) = \sum_{k=-\infty}^\infty  (k^2 - \nu k^4)\Theta(\abs{k}-k_c) 
    v_k  e^{2\pi i kx /L}\,, 
\end{align}
where $\Theta(\abs{k}-k_c)$ is the Heaviside theta function which only assumes a nonzero value when $\abs{k}\geq k_c$.
This operator only has nonpositive spectra, so is consistent with the form required by the UT algorithm.
In fact, writing $T_{diss} = -\mathcal{L}^v\mathcal{L}^v_* $ we may even write down $\mathcal{L}^v$ by calculating its square root using the Fourier basis  
\begin{align}
    \mathcal{L}^vv(x) = \sum_{k=-\infty}^\infty  \sqrt{k^2 - \nu k^4} \Theta(\abs{k}-k_c) v_k e^{2\pi i kx /L} \,.
\end{align}
Notice that $\mathcal{L}^v=\mathcal{L}^v_*$ is self adjoint. 
By using $T_{diss}$ and $\mathcal{L}^v$ as new differential operators for our system, we may extract a thermodynamically consistent reduction of the KS equation
\begin{align}
    \frac{\partial v}{\partial t} + vv_x &= T_{diss}v \,,\label{eq:vdiss} \\ \label{eq:sigmadiss}
    \frac{\partial\sigma}{\partial t}+\partial_x(\sigma v) &= \frac1T (\mathcal{L}^vv)^2 \,.
\end{align}
Given that we are mainly interested in the development of the global entropy $S = \int_0^L \sigma dx$, we integrate \eqref{eq:sigmadiss} over the spatial domain to find
\begin{align}
    \frac{\partial S}{\partial t} = \int_0^L \frac1T(\mathcal{L}^vv)^2 dx \,, \label{eq:sdiss} 
\end{align}
which is the rate of total entropy production.
\vspace{\baselineskip}\newline
\noindent We conclude this section with a diversion to an alternative perspective on how the KS equation conflicts with the metriplectic viewpoint.
It is possible to obtain a symmetric bracket which recovers the behavior of the KS system.
To do this, we may the Kulkarni-Nomizu product of the bivectors
\begin{align}
    M(dF,dG) = F_\sigma G_\sigma
\end{align}
and \begin{align}
    \Sigma(dF,dG) = - \frac{\partial F_{v}}{\partial x} \frac{\partial G_v}{\partial x} + \nu \frac{\partial^2 F_v}{\partial x^2}\frac{\partial ^2 G_v}{\partial x^2} \,,
\end{align}
which leads to equations for the velocity and entropy
\begin{align}
    \frac{\partial v}{\partial t}+vv_x &= -v_{xx}-\nu v_{xxxx} \,, \\ \nonumber
    \frac{\partial \sigma}{\partial t}+\partial_x(\sigma v) &= - \frac1{T}(v_x)^2+\frac{\nu}{T}(v_{xx})^2 \,.
\end{align}
This manages to recover the KS equation, but notice how the total entropy now behaves as
\begin{align}
    \frac{\partial S}{\partial t} = \frac1T \int_0^L \nu (v_{xx})^2 - (v_x)^2 dx \,,
\end{align}
which is no longer strictly nonnegative.
This owes to the fact that the bivector $\Sigma$ is no longer positive definite, as is required for a system which produces entropy.
\section{Numerical Study of Dissipative KS System}

\begin{figure*}
    \begin{subfigure}{0.4\linewidth}
        \includegraphics[width=\linewidth]{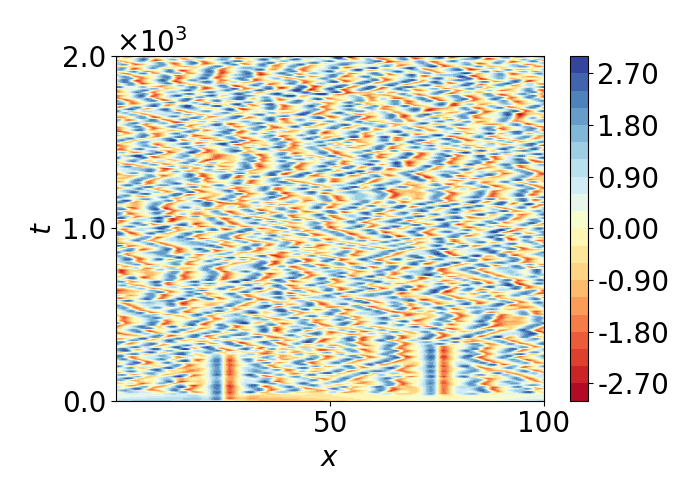}
        \caption{}
        \label{fig:kt05}
    \end{subfigure}
    \hfill
    \begin{subfigure}{0.4\linewidth}
        \includegraphics[width=\linewidth]{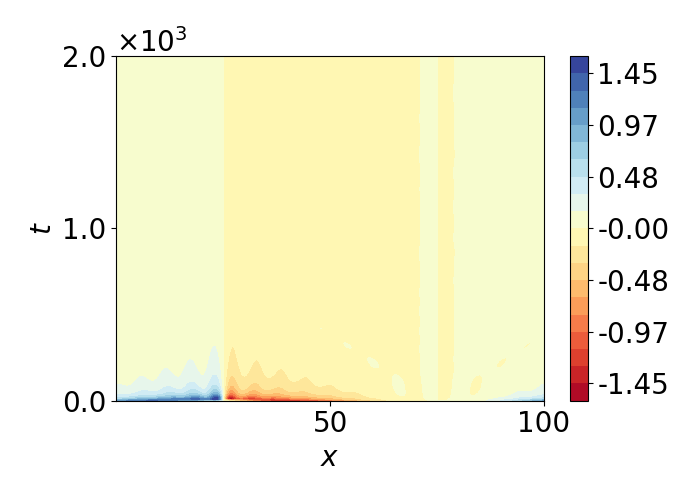}
        \caption{}
        \label{fig:kt05met}
    \end{subfigure}

    \begin{subfigure}{0.4\linewidth}
        \includegraphics[width=\linewidth]{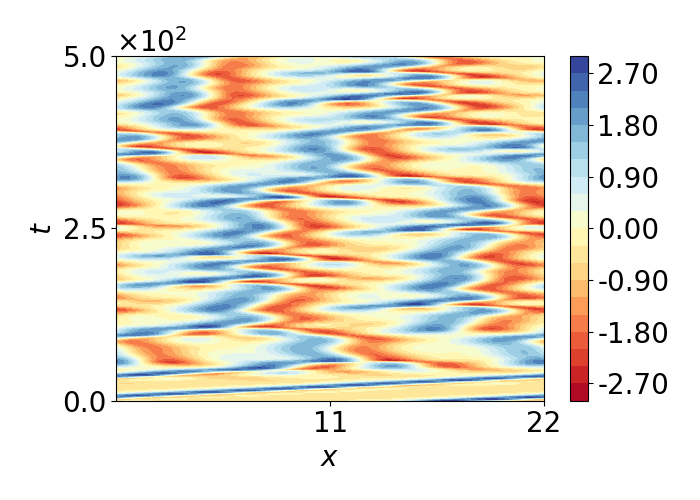}
        \caption{}
        \label{fig:cde10a}
    \end{subfigure}
    \hfill
    \begin{subfigure}{0.4\linewidth}
        \includegraphics[width=\linewidth]{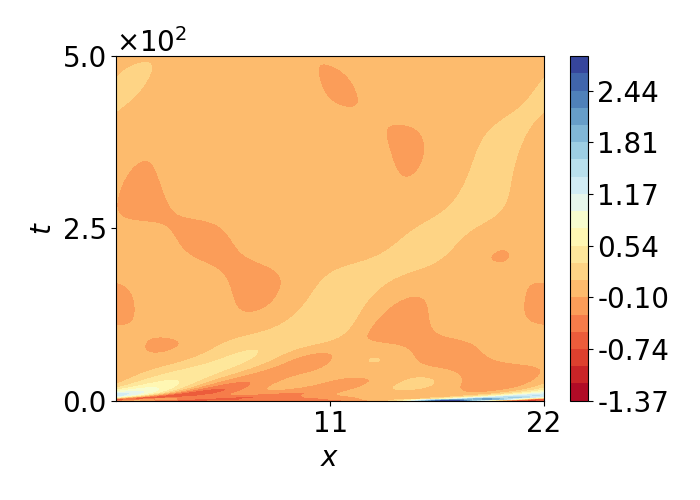}
        \caption{}
        \label{fig:cde10amet}
    \end{subfigure}

    \begin{subfigure}{0.4\linewidth}
        \includegraphics[width=\linewidth]{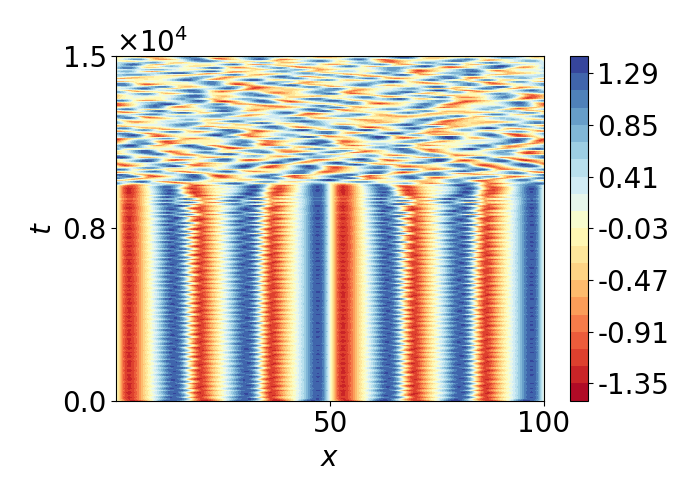}
        \caption{}
        \label{fig:hnz86b}
    \end{subfigure}
    \hfill
    \begin{subfigure}{0.4\linewidth}
        \includegraphics[width=\linewidth]{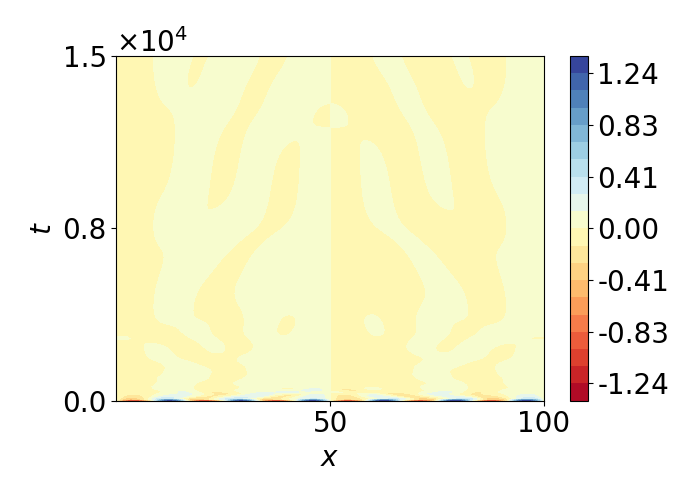}
        \caption{}
        \label{fig:hnz86bmet}
    \end{subfigure}

    \begin{subfigure}{0.4\linewidth}
        \includegraphics[width=\linewidth]{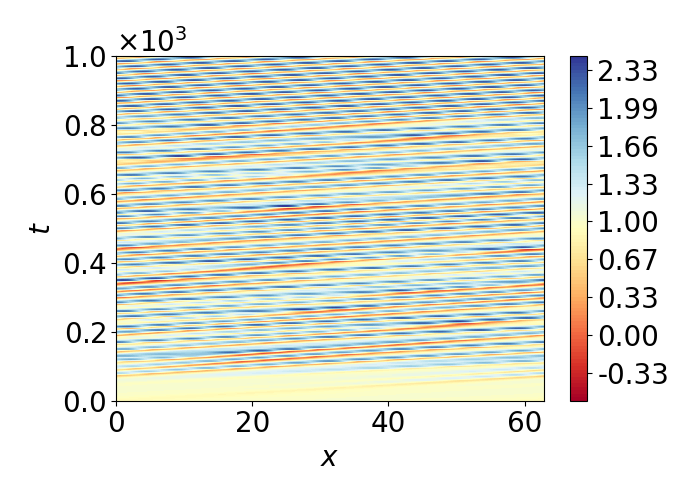}
        \caption{}
        \label{fig:sech}
    \end{subfigure}
    \hfill 
    \begin{subfigure}{0.4\linewidth}
        \includegraphics[width=\linewidth]{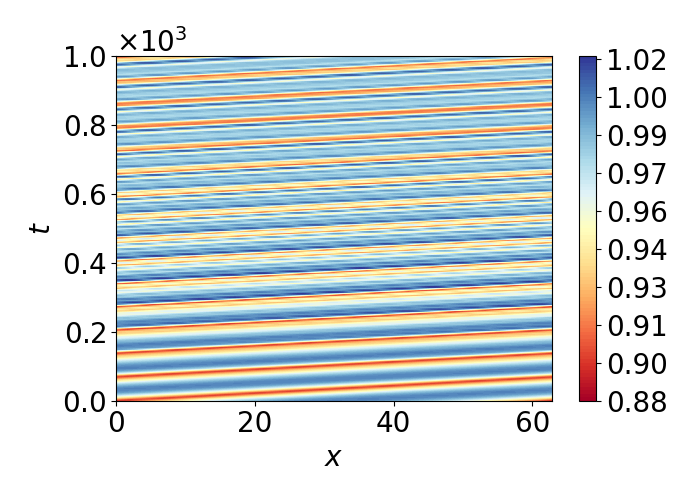}
        \caption{}        
        \label{fig:sechmet}
    \end{subfigure}
    \caption{Solutions of the KS equation (left) are dissipated under metriplectic dynamics (right). Each row corresponds to a different initial condition (see text for details).}
    \label{fig:metdiss}
\end{figure*}
\begin{figure*}[t]
\centering
    \begin{subfigure}{0.4\linewidth}
        \includegraphics[width=\linewidth]{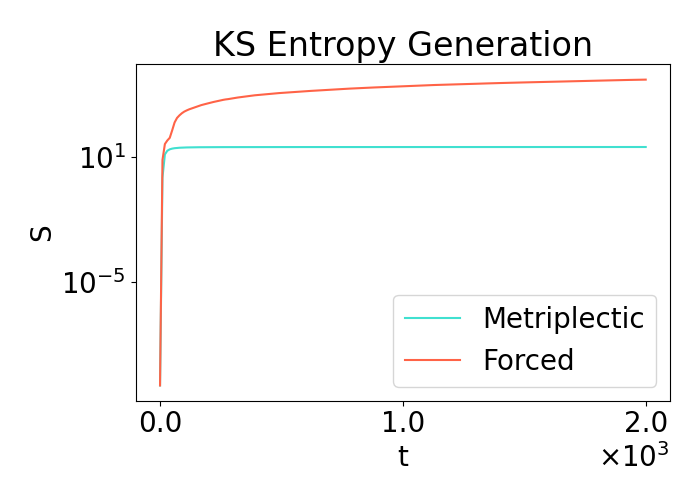}
        \caption{$\cos(\frac{x}{16})(1+\sin(\frac{x}{16}))$}
    \end{subfigure}
    \hfill
    \begin{subfigure}{0.4\linewidth}
        \includegraphics[width=\linewidth]{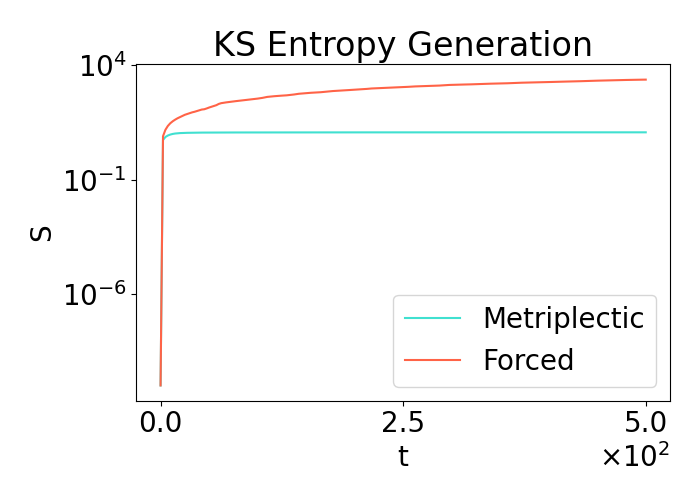}
        \caption{Traveling wave}
    \end{subfigure}

    \begin{subfigure}{0.4\linewidth}
        \includegraphics[width=\linewidth]{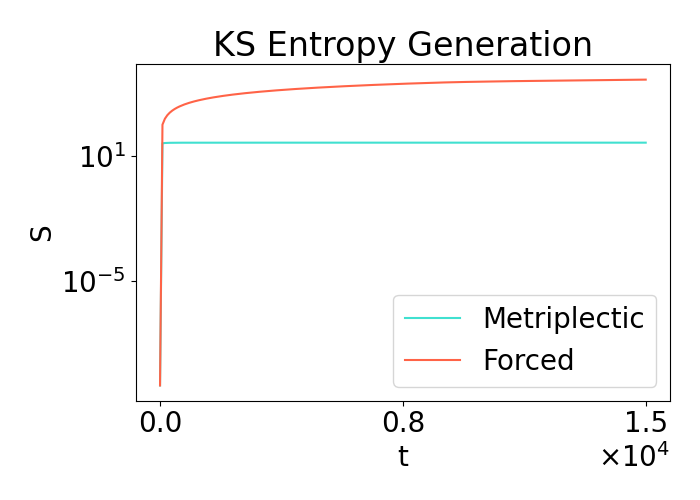}
        \caption{Strange hyperbolic point}
    \end{subfigure}
    \hfill 
    \begin{subfigure}{0.4\linewidth}
        \includegraphics[width=\linewidth]{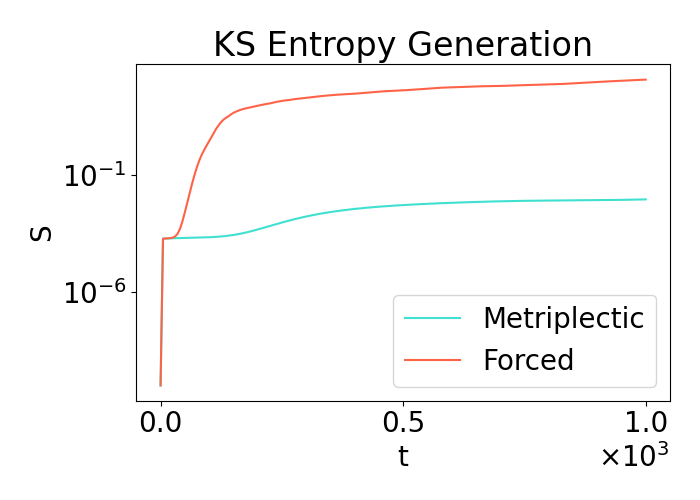}
        \caption{Sech profile}
    \end{subfigure}
    \caption{Entropy generation associated with metriplectic dynamics of the Kassam Trefethen benchmarking conditions, traveling wave solutions, a strange hyperbolic point, and a hyperbolic secant profile.}
    \label{fig:entropies}
\end{figure*}

\begin{figure}[h!]
    \centering
    \includegraphics[width=0.55\linewidth]{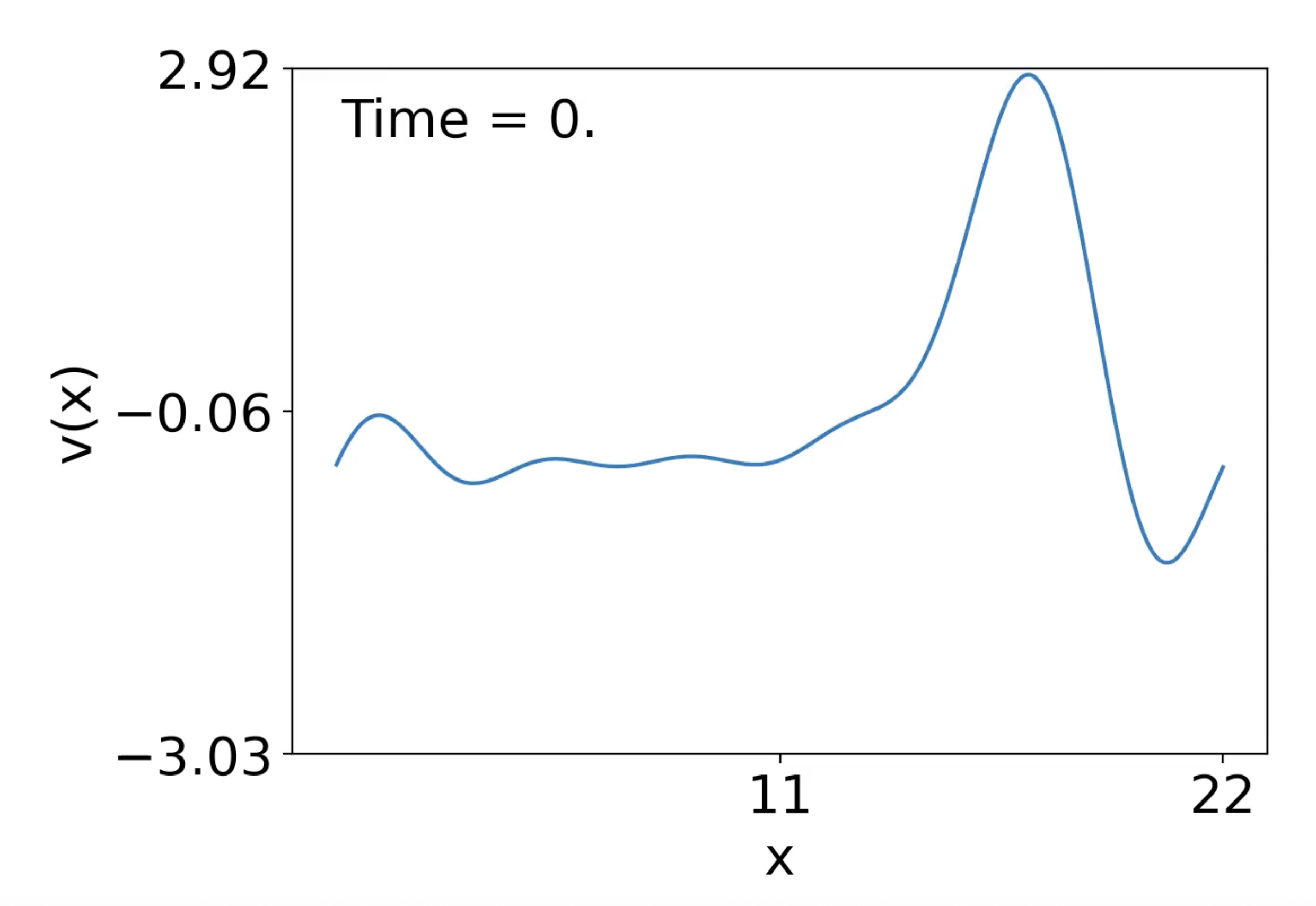}
    \caption{Interpolated profile of TW1 used for initialization.}
    \label{fig:tw1prof}
\end{figure}

\noindent With the thermodynamically consistent KS equation that we have defined, we now perform a numerical study to determine how solutions behave under a thermodynamically consistent system.
The definition of $T_{diss}$ and $\mathcal{L}$ using the Fourier basis make a pseudospectral spatial discretization especially convenient.
The $vv_x$ nonlinearity convolution is computed pseudospectrally using the 3/2 dealiasing rule and FFTs are computed using the SciPy Python package.\cite{Canuto2006}
Due to the stiffness of the KS equation, we use the fourth order exponential time integrator developed by Cox and Matthews and used by Kassam and Trefethen and Cvitanovic et al.\ for studies of the KS equation.\cite{CoxMatthews,KassamTrefethen,CDE10}
The local truncation error for an exponential method does not scale with stiff eigenvalues, so this method performs well compared to similar methods. 
\vspace{\baselineskip}\newline
\noindent 
To assess the effects of thermodynamic consistency, we sample a variety of initial conditions to studies of the KS equation that have been previously considered in the literature.
These conditions show initial wavelike behavior and interactions before disintegrating into chaos and turbulence.
By this, we mean that the motion qualitatively forms large amplitude oscillation in time and space with no clear period.
We closely consider the benchmarking initial condition used by Kassam and Trefethen, \cite{KassamTrefethen}
\begin{align}
    v(x,0) &= \cos(\frac{x}{16})(1+\sin(\frac{x}{16}))\,,  \\ \nonumber\hspace{2em} 0&\leq x \leq 32\pi \,,\, \nu = 1\label{eq:kt} \,.
\end{align}
Another established wave that is absorbed into chaos and a strange attractor is a strange hyperbolic point observed by Hyman, Nicolaenko, and Zareski given by initial profile \cite{HNZOrder}
\begin{align}
    v(x,0) &= \frac{-21\sin(12\pi x/\sqrt{253}) -\sin(2\pi x/\sqrt{253})}{\sqrt{253}} \,, \\ \nonumber
    0&\leq x\leq 2\pi\sqrt{253}\,,\, \nu = 4 \,.
\end{align}
We also consider relative equilibrium solutions through a perturbed traveling wave found by Cvitanovic et al. and a hyperbolic secant profile which evolves to a traveling wave.
The initial condition for the perturbed traveling wave is initialized by interpolating the profile given in Cvitanovic et al.;\cite{CDE10} we display this profile in Figure \ref{fig:tw1prof}.
Such solutions are useful to understand what behavior exhibited by the KS equation can occur in a metriplectic setting.
\vspace{\baselineskip}\newline
\noindent We survey the evolution of these cases under the metriplectic system throughout Figure \ref{fig:metdiss}.
The benchmarking conditions of Kassam and Trefethen shown in Figure \ref{fig:kt05} start to nonlinearly beat and form higher order waves, only to be depleted once large gradients form in \ref{fig:kt05met}.
From that stage, the motion is confined to small wiggles as the profile shrinks to zero.
A perturbed traveling wave solution to the KS equation, which originally circles the domain twice before falling apart in Figure \ref{fig:cde10a}, slows tremendously and loses its energy in a similar way in \ref{fig:cde10amet}.
Late transitions to chaos which characterize strange hyperbolic points as shown in \ref{fig:hnz86b} never occur, as the motion has been depleted long before crisis in \ref{fig:hnz86bmet}.
Each of these previously complex solutions decreases to zero over the course of the metriplectic dynamics.
The hyperbolic secant profile is slightly different from these cases.
Under the dynamics of the KS equation, this profile undergoes chaos and oscillates rapidly until it forms a period four traveling wave in Figure \ref{fig:sech}.
When the forcing is eliminated in the metriplectic system, the profile instead circles the domain while undergoing nonlinear beating to a higher order wave in \ref{fig:sechmet}.
This mode is not fully depleted, and has not achieved a constant value. 
However, we conclude from these examples that the metriplectic system dynamics reduces the amplitude of oscillations of the KS equation.
\vspace{\baselineskip}\newline
\noindent We perform some analysis to characterize the equilibria and possibilities for traveling waves in the metriplectic KS system we've defined.
Through the method of Lagrange multipliers, equilibria of the metriplectic system that maximize the entropy must satisfy \cite{MetRelaxEq}
\begin{align}
    \frac{\delta S}{\delta v}-\lambda \frac{\delta H}{\delta v} = -\lambda v& = 0 \,, \\ \nonumber
    \frac{\delta S}{\delta \sigma } - \lambda \frac{\delta H}{\delta \sigma} = 1-\lambda T&=0 \,, 
\end{align}
showing that maximum entropy solutions satisfy $v = 0$.
Galilean invariance of the KS equation would suggest that this also allows constant $v$ solutions to have maximum entropy.\cite{CDE10}
However, equilibria are not required to satisfy this variational principle, and the hyperbolic secant profile mentioned above does not appear to.
\vspace{\baselineskip}\newline
\noindent To characterize all equilibria of the KS equation, we consider the deviations of solutions to the KS equation with the spatial average value
\begin{align}
    \bar{v} = \frac1L\int_0^L v(x,t) dx \,.
\end{align}
By integrating the KS equation over space, we have
\begin{align}
    \frac{\partial }{\partial t}\int_0^L v dx
    + \int_0^L v v_x dx = \int_0^L -\mathcal{L}^v\mathcal{L}^v_* v dx \,,
\end{align}
or
\begin{align}
    L \frac{\partial \bar{v}}{\partial t} = -\int_0^L  \mathcal{L}^v\mathcal{L}^v_* v dx - \bigg[\frac12 v^2 \bigg]_0^L = 0 \,,
\end{align}
as $\mathcal{L}^v\mathcal{L}^v_*$ only acts on high frequency modes which are zero on averaging.
This establishes that $\bar{v}$ is constant in space and time.
As a result, we may rewrite the KS equation as 
\begin{align}
    \frac{\partial(v -\bar{v})}{\partial t} &+ (v-\bar{v})\partial_x(v-\bar{v})+\bar{v}\partial_x(v-\bar{v}) \\ \nonumber
    &= -\mathcal{L}^v\mathcal{L}^v_*(v-\bar{v}) \,.
\end{align}
Multiplying by $v-\bar{v}$ and integrating over space yields
\begin{align}
    \frac{\partial}{\partial t}\int_0^L (v&-\bar{v})^2 dx 
    =-\int_0^L (v-\bar{v})\mathcal{L}^v\mathcal{L}^v_*(v-\bar{v}) dx \\ \nonumber &-\int_0^L ((v-\bar{v})^2 + \bar{v}(v-\bar{v}))\partial_x(v-\bar{v})dx  \\ \nonumber
    &= -\int_0^L (\mathcal{L}^v_*(v-\bar{v}))^2 dx \leq 0 \,,
\end{align}
where we have integrated by parts and used the fact that the second integral vanishes.
With the conservation of $\bar{v}$, this result implies that the metriplectic KS system evolves arbitrary initial conditions to their average value.
For a solution to be an equilibrium or periodic in time, the inequality must be saturated with $\mathcal{L}^v_*(v-\bar{v}) = 0$ for all times.
But then an equilibrium satisfies
\begin{align}
    vv_x = -\mathcal{L}^v\mathcal{L}^v_* v = 0 \,,
\end{align}
and integrating in space yields
\begin{align}
    \frac12 v(x)^2 = \frac12 v(0)^2 \,.
\end{align}
Thus, we conclude that the only equilibria of the metriplectic KS system are constants.
\vspace{\baselineskip}\newline
\noindent We now show the entropy associated with the metriplectic dynamics in the cases considered in Figure \ref{fig:entropies}.
Having defined the total entropy evolution in terms of  $v(x,t)$, we may allow $v$ to be determined by the original KS equation and produce a positive definite entropy.
Our main result focuses on the distinction between this forced system entropy and purely metriplectic dynamics.
As required by our method, the metriplectic dynamics show positive definite entropy growth.
But we see that in every case, the dynamics of the original KS equation cause the entropy to increase further than the dissipative dynamics alone.
In contrast to our expectations, the forcing effects which cause nontrivial patterns to emerge in the KS equation do not decrease the metriplectic entropy.
This idea establishes the consistency of pattern formation with the second law of thermodynamics.

\section{Variation of Forcing Strength}

\begin{figure*}
    \centering
    \begin{subfigure}{0.4\textwidth}
        \includegraphics[width=\linewidth]{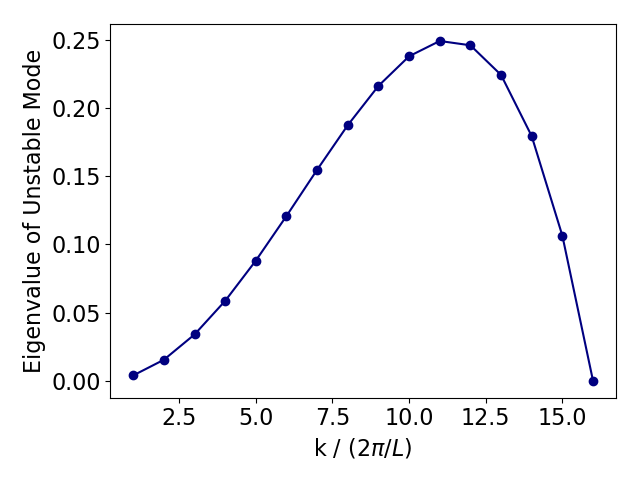}
        \caption{Spectra for $\nu = 1$ and $L = 32\pi$.}
        \label{fig:ktspec}
    \end{subfigure}
    \hfill
    \begin{subfigure}{0.4\textwidth}
        \includegraphics[width=\linewidth]{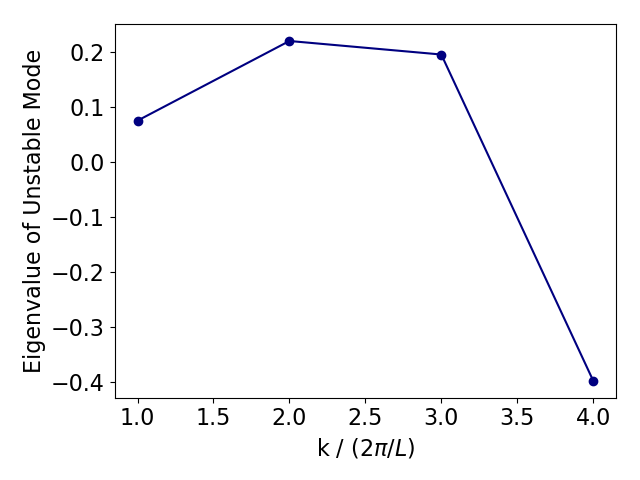}
        \caption{Spectra for $\nu = 1$ and $L = 22$.}
        \label{fig:tw1spec}
    \end{subfigure}    
    \caption{Positive spectra of the Kassam Trefethen benchmarking and Cvitanovic et al TW1 condition parameters.}
    \label{fig:travwave}
\end{figure*}
\begin{figure*}[t!]
    \begin{subfigure}{0.4\textwidth}
        \centering
        \includegraphics[width=\linewidth]{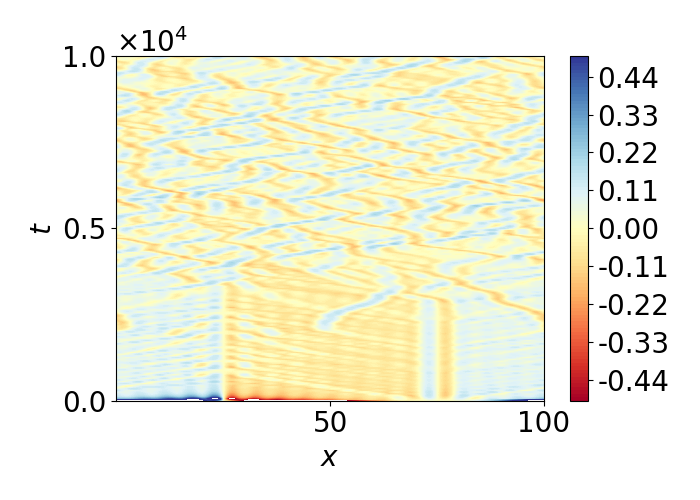}
        \caption{$\epsilon = 0.01$}
        \label{fig:ktepsupa}
    \end{subfigure}
    \hspace{1in}
    \begin{subfigure}{0.4\textwidth}
        \centering
        \includegraphics[width=\linewidth]{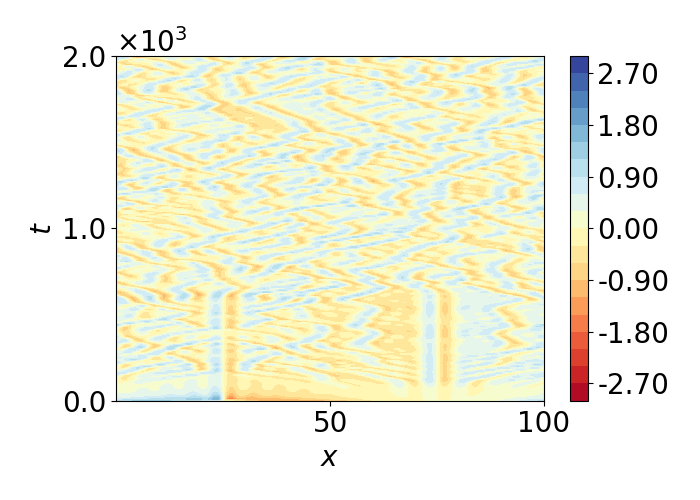}
        \caption{$\epsilon = 0.2$}
        \label{fig:ktepsupb}
    \end{subfigure}

    \begin{subfigure}{0.4\textwidth}
        \centering
        \includegraphics[width=\linewidth]{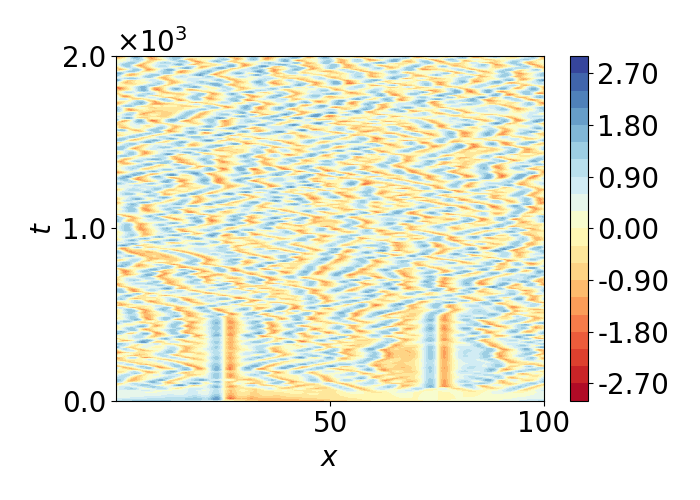}
        \caption{$\epsilon = 0.5$}
        \label{fig:ktepsupc}
    \end{subfigure}
    \hspace{1in}
    \begin{subfigure}{0.4\textwidth}
        \centering
        \includegraphics[width=\linewidth]{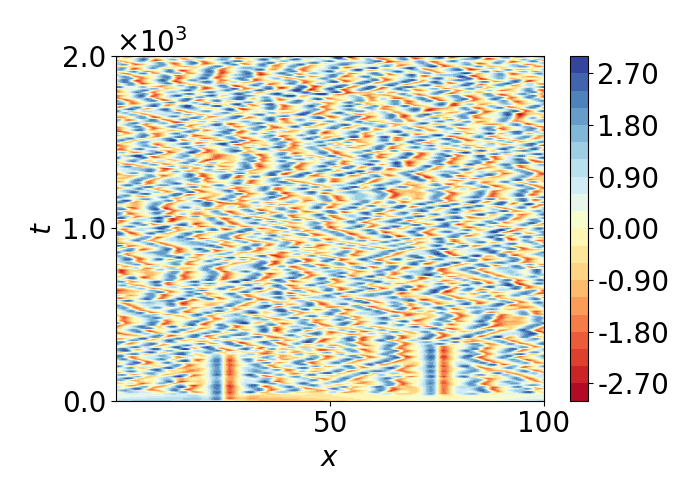}
        \caption{$\epsilon = 1$}
        \label{fig:ktepsupd}
    \end{subfigure}
    \caption{Spatiotemporal variation of $v(x,t)$ as the growth rate of positive spectra of  $T_{KS}$ is increased linearly. As the parameter $\epsilon$ decreases, the transition to chaos remains, but occurs later in the dynamics and the amplitude of the oscillations decreases.}
    \label{fig:ktepsup}
\end{figure*}
\begin{figure*}[t!]
    \centering
    \begin{subfigure}[b]{0.3\textwidth}
        \centering
        \includegraphics[width=\linewidth]{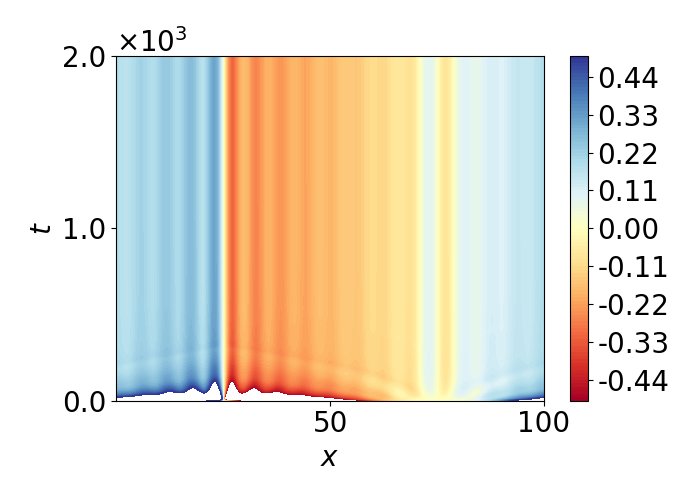}
        \caption{$n=1$}
        \label{fig:ktcontourv1}
    \end{subfigure}
    \hfill
    \begin{subfigure}[b]{0.3\textwidth}
        \centering
        \includegraphics[width=\linewidth]{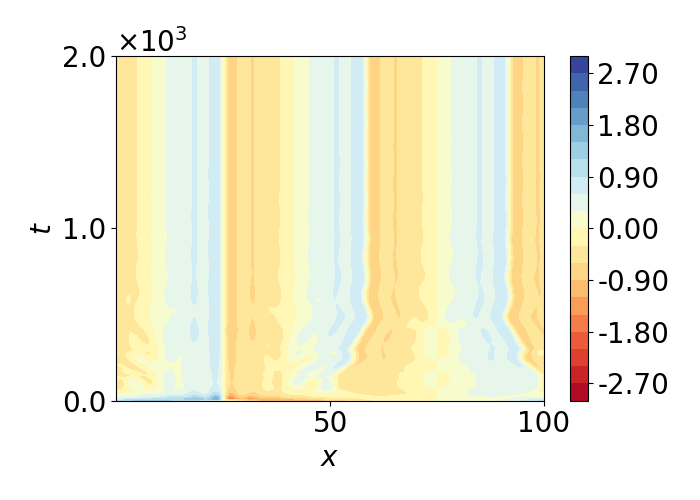}
        \caption{$n=3$}
        \label{fig:ktcontourv3}
    \end{subfigure} 
    \hfill
    \begin{subfigure}[b]{0.3\textwidth}
        \centering
        \includegraphics[width=\linewidth]{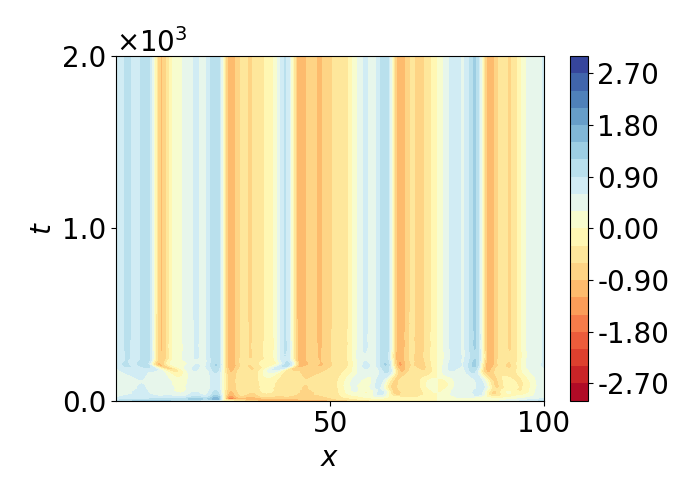}
        \caption{$n=5$}
        \label{fig:ktcontourv5}
    \end{subfigure}

    \begin{subfigure}[b]{0.3\textwidth}
        \centering
        \includegraphics[width=\linewidth]{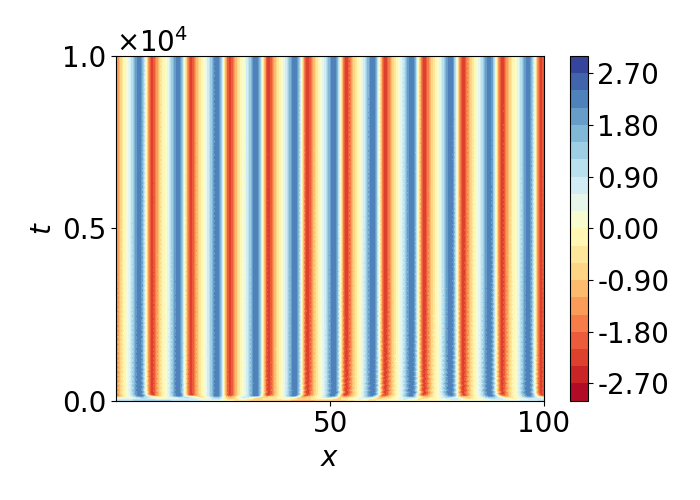}
        \caption{$n = 11$}
        \label{fig:ktcontour11}
    \end{subfigure}
    \hfill
    \begin{subfigure}[b]{0.3\textwidth}
        \centering
        \includegraphics[width=\linewidth]{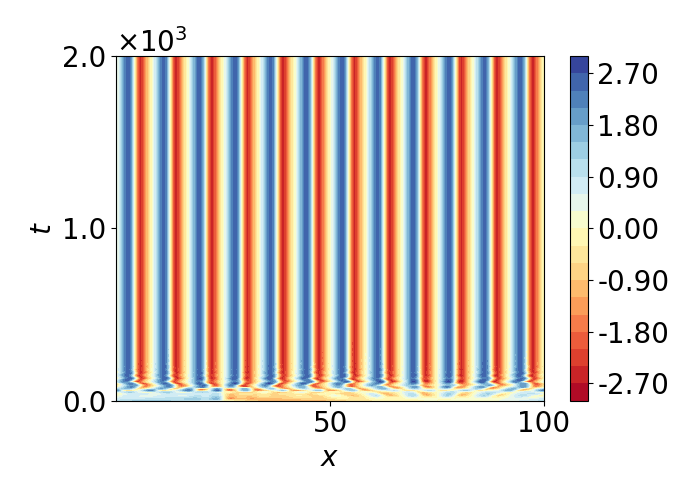}
        \caption{$n= 12,13$}
        \label{fig:ktcontour1213}
    \end{subfigure}
    \caption{Contours of $v(x,t)$ equilibria which form from the injection of one or more labeled driven positive spectra into the metriplectic KS system.}
    \label{fig:kteq}
\end{figure*}

\begin{figure*}[t!]
    \centering
    \begin{subfigure}[b]{0.3\textwidth}
        \centering
        \includegraphics[width=\linewidth]{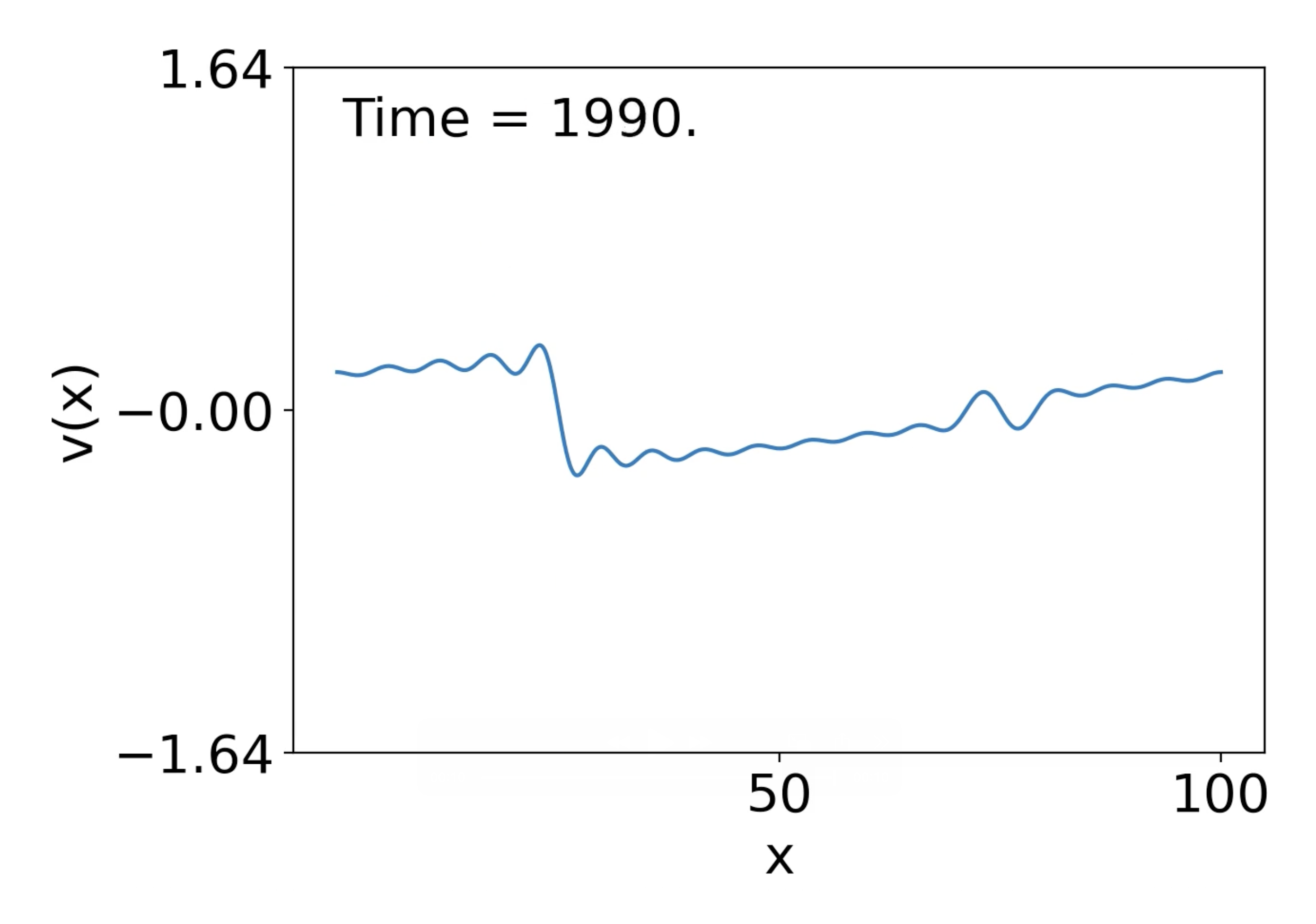}
        \caption{$n = 1$}
        \label{fig:ktv1}
    \end{subfigure}
    \hfill
    \begin{subfigure}[b]{0.3\textwidth}
        \centering
        \includegraphics[width=\linewidth]{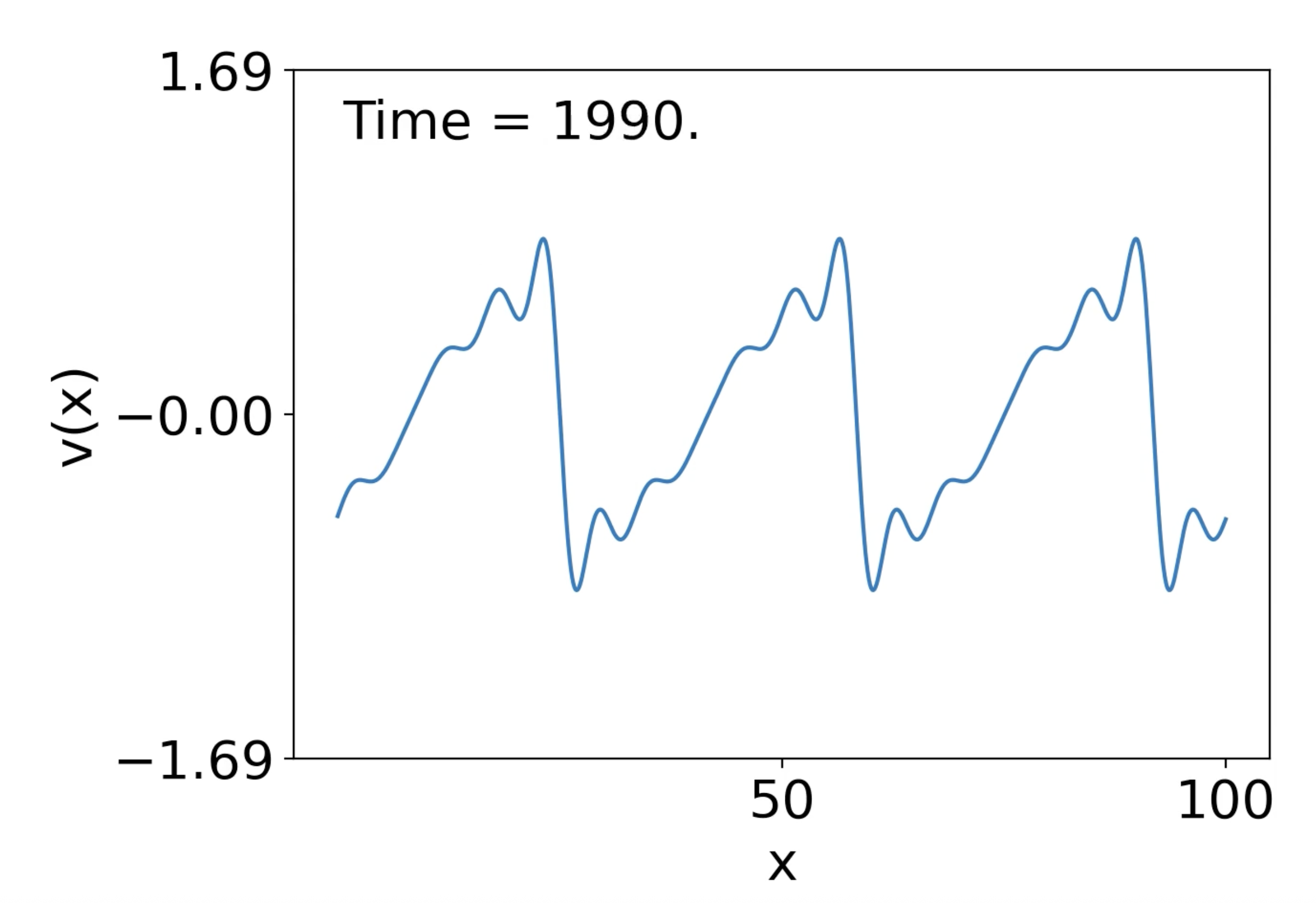}
        \caption{$n = 3$ }
        \label{fig:ktv3}
    \end{subfigure}
    \hfill
    \begin{subfigure}[b]{0.3\textwidth}
        \centering
        \includegraphics[width=\linewidth]{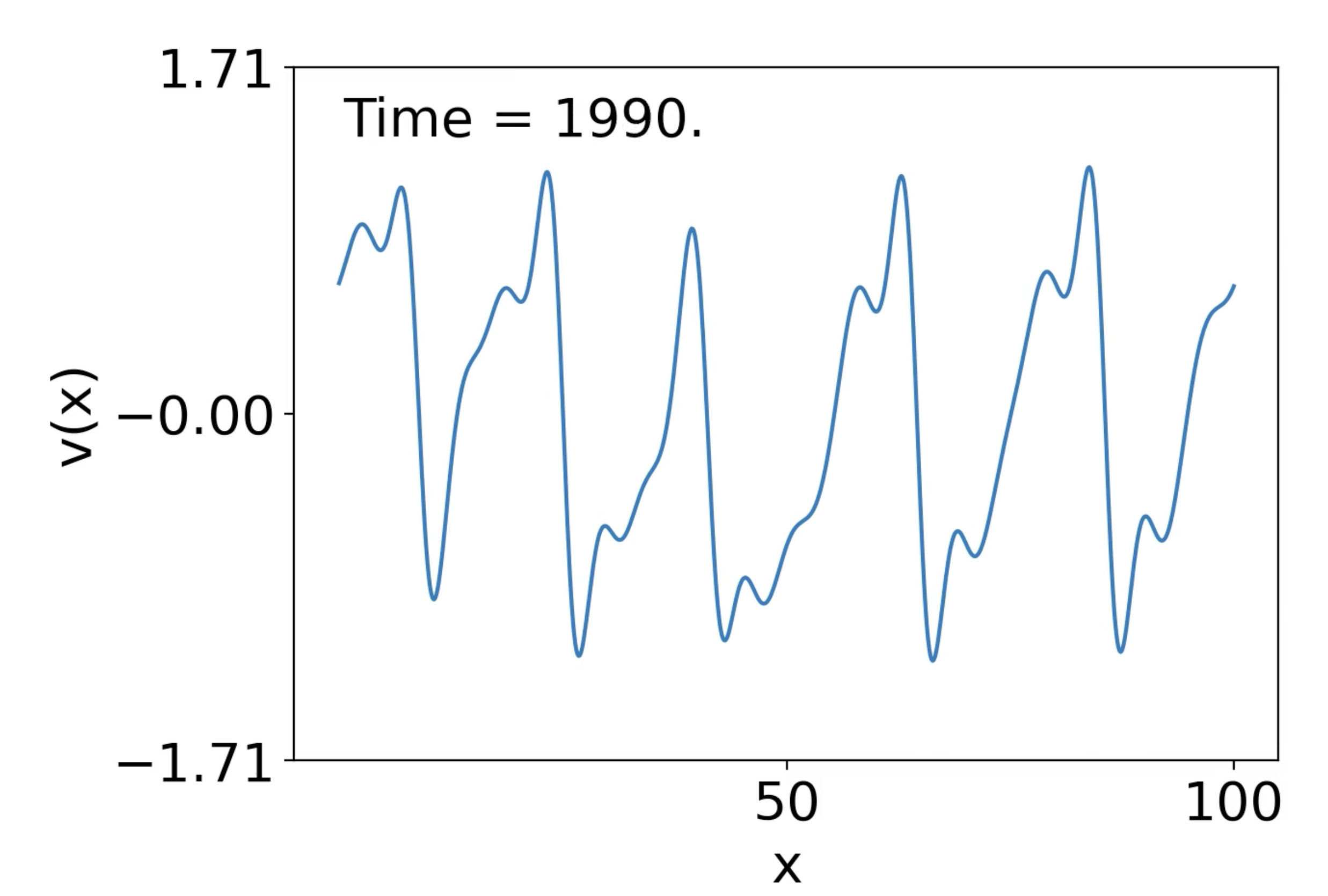}
        \caption{$n = 5$}
        \label{fig:ktv5}
    \end{subfigure}
    
    \caption{Profiles of equilibria found with $n=1,3,5$ driven spectra.}
    \label{fig:ktsmallnprofs}
\end{figure*}
\begin{figure*}[t!]
    \begin{subfigure}[b]{0.3\textwidth}
        \centering
        \includegraphics[width=\linewidth]{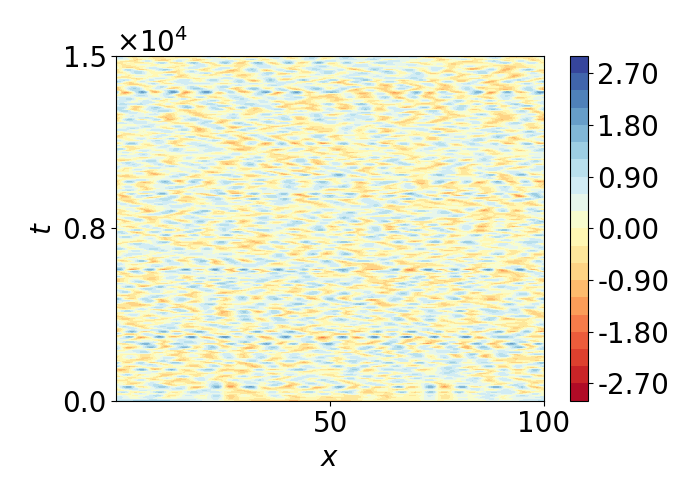}
        \caption{$n=14$}
        \label{fig:kt14}
    \end{subfigure}
    \hfill 
    \begin{subfigure}[b]{0.3\textwidth}
        \centering
        \includegraphics[width=\linewidth]{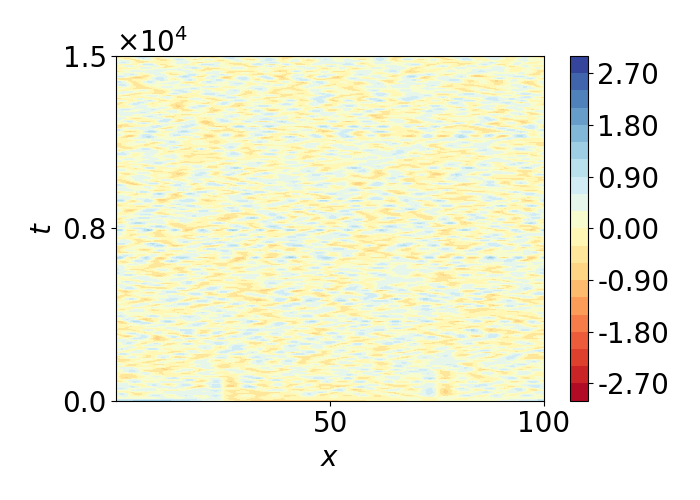}
        \caption{$n=15$ }
        \label{fig:kt15}
    \end{subfigure}
    \hfill
    \begin{subfigure}[b]{0.3\textwidth}
        \centering
        \includegraphics[width=\linewidth]{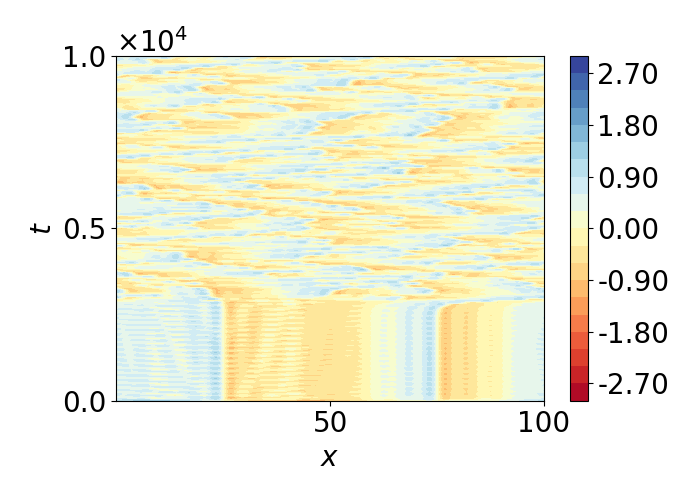}
        \caption{$n = 1,2,3,4$}
        \label{fig:ktleft4}
    \end{subfigure}
    
    \caption{Chaotic solutions of unstable wavenumber injection.}
    \label{fig:ktchaos}
\end{figure*}

 \begin{figure*}[t!]
    \begin{subfigure}{0.4\textwidth}
        \centering
        \includegraphics[width=\linewidth]{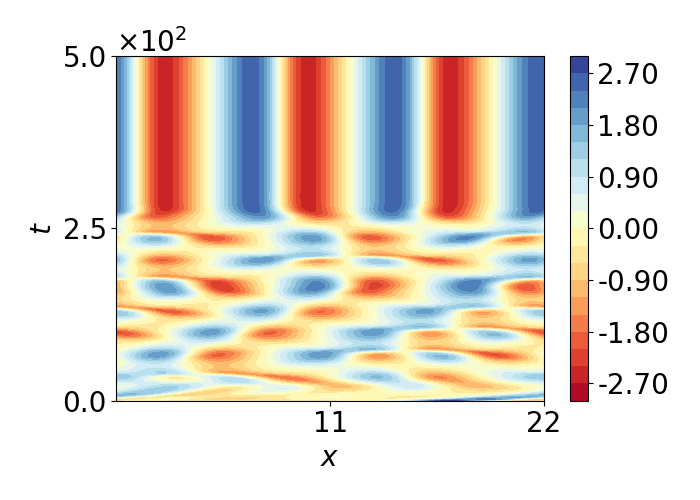}
        \caption{$n = 3$}
        \label{fig:cdemaxeq}
    \end{subfigure}
    \hspace{1in}
    \begin{subfigure}{0.4\textwidth}
        \centering
        \includegraphics[width=\linewidth]{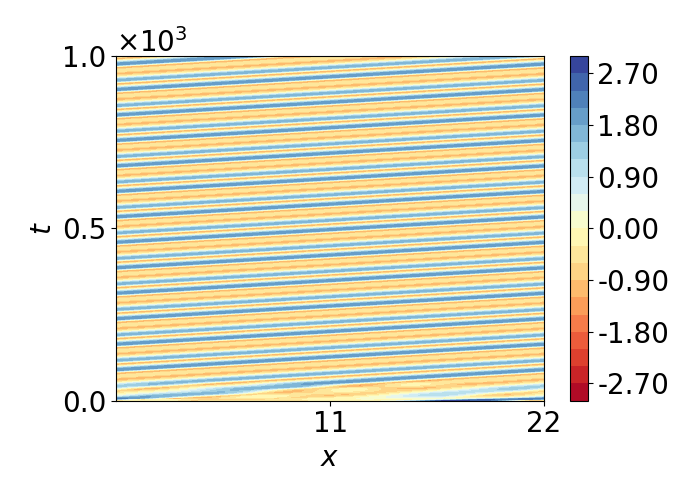}
        \caption{$n = 1$ and $n =3$}
        \label{fig:cdepersist}
    \end{subfigure}

    \begin{subfigure}{0.4\textwidth}
        \centering
        \includegraphics[width=\linewidth]{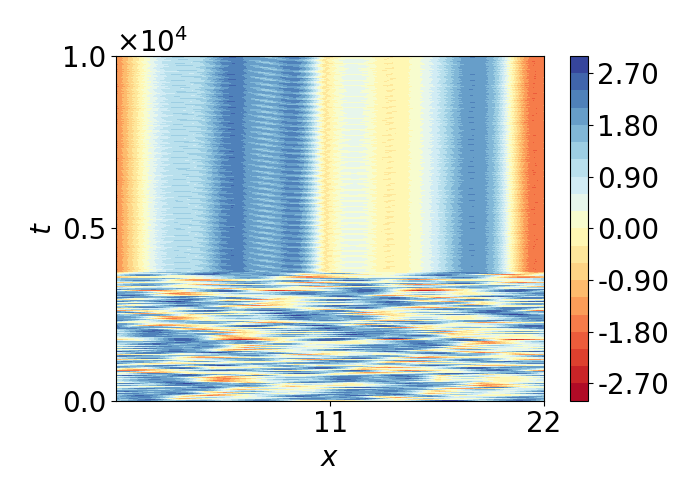}
        \caption{$n = 2$ and $n = 3$}
        \label{fig:cdetransition}
    \end{subfigure}
    \hspace{1in}
    \begin{subfigure}{0.4\textwidth}
        \centering
        \includegraphics[width=\linewidth]{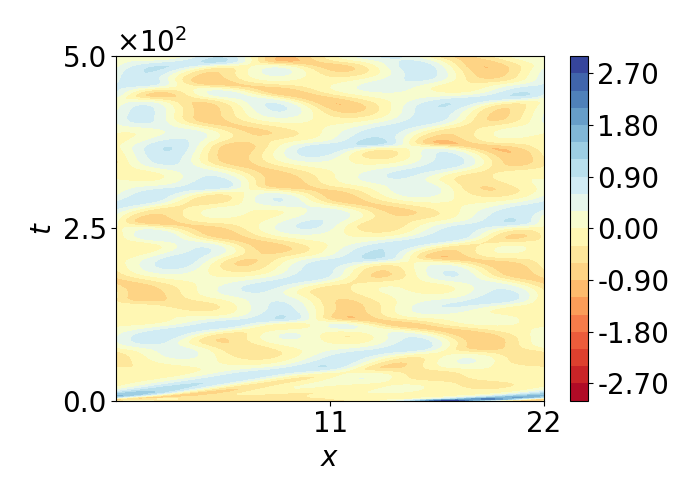}
        \caption{$\epsilon = 0.2$ }
        \label{fig:cde5}
    \end{subfigure}
    \caption{Equilibria for the traveling wave solution TW1.}
    \label{fig:forcingcde10}
\end{figure*}
\noindent 

The analysis of the KS equation and its metriplectic reduction conducted in the previous section motivates study of the instabilities of the KS equation as an external disturbance to a metriplectic system.
Now we will analyze the effect of the positive spectra on two sets of initial conditions previously mentioned.
These spectra were all removed in the definition of the metriplectic KS system from previous sections.
In order to study the addition of energy injection, we can consider two approaches. We rewrite $T_{KS}$ as the sum of dissipative and driving Fourier components
\begin{align}
    T_{KS} = -\mathcal{L}^v\mathcal{L}^v_* + \epsilon
     (T_{KS} + \mathcal{L}^v\mathcal{L}^v_*) \,,
\end{align}
where $T_{KS}+\mathcal{L}\mathcal{L}_*$ represents the positive spectra of the KS antidiffusion and hyperviscosity, shown for example in Figure \ref{fig:travwave}.
One option for reintroducing instability to the metriplectic KS system is to slowly increase $\epsilon$, allowing all positive Fourier coefficients at once albeit with smaller growth rate.
An alternative to this approach would be to drive individual small wavenumbers as subsets of the positive spectra. 
Choosing modes to allow into the KS equation allows the study of which combinations of drivers allow certain behavior in the KS system.
\vspace{\baselineskip}\newline 
The approach of adding forcing components to a Hamiltonian or metriplectic system can be compared with the notion of a port-Hamiltonian (pH) system.
A pH system describes the interactions of an otherwise Hamiltonian system with its environment.\cite{2002VDSMaschke}
This is accomplished by means of a Dirac structure and additional pairs of power variables to describe the system boundaries, and describes energy flow through a boundary, external impacts of the Hamiltonian dynamics, and energy dissipation.
Irreversible pH systems are also able to account for entropy production by introducing a contact vector field.\cite{Eberard2007}
This is similar to our approach in that it generates entropy and conserves the total energy of the system.
However, while the positive spectra in our work correspond to external forcing, they are not interpreted as passing through a spatial boundary of the system.
\vspace{\baselineskip}\newline
We perform a numerical study on the Kassam Trefethen initial conditions.
As studied there, this initial condition is defined with 15 wavenumbers with positive spectra given the associated parameters $L = 32\pi $ and $\nu = 1$, shown in Figure \ref{fig:ktspec}.
This allows many subsets of modes to consider to determine how this initial condition transitions to chaos.
Under the dynamics of the unmodified KS equation, \eqref{eq:kt} evolves into an ordered pattern of moving waves and transitions into a chaotic attractor at later times.
However, the metriplectic system depletes the energy of this mode and only small ripples emerge before they fade away.
Consequently, forcing of this system is necessary for nontrivial patterns to be sustained.
\vspace{\baselineskip}\newline
We begin by increasing the growth rate of all positive spectra at once.
This process is illustrated in Figure \ref{fig:ktepsup}; the dynamics of the original KS equation are shown in Figure \ref{fig:ktepsupd} with $\epsilon=1$.
By decreasing $\epsilon$ to half the forcing strength of the original KS equation, we see similar patterns as the motion progresses and the breakdown into the strange attractor from the KS equation.
However, we notice that the transition to nonlinear dynamics occurs later.
This pattern continues as $\epsilon$ decreases to $0.2$ and $0.01$, suggesting that the presence of all positive spectra is responsible for an eventual transition to chaos regardless of the size of the growth rates.
\vspace{\baselineskip}\newline
With this analysis of what occurs when all mode forcings are controlled simultaneously, we consider the effects of injecting individual wavenumbers into the dynamics throughout Figure \ref{fig:kteq}.
We begin by adding the weakest mode corresponding to the first $n = 1$ nonzero wavenumber at which the spectrum assumes a positive value just above zero.
At this value, the dynamics are dissipated but achieve a steady state shown in Figures \ref{fig:ktcontourv1} and \ref{fig:ktv1}.
This pattern continues for the third and fifth positive spectra $n=3$ and $n = 5$, where the amplitude of the steady state increases alongside the time to achieve it.
Equilibria also form when the most unstable modes are driven.
When $n = 11$, the mode appears to shake, but the dynamics continue to have frequency 11 for long time scales.
A combination of modes $n = 12$ and $n = 13$ also forms an equilibrium.
The structure of the resulting steady state also follows the period of the driven wavenumber, though small fluctuations appear.
In the last case, the periodicity of the equilibrium follows the $n= 12$ mode; we might understand this as the stronger mode setting the dynamics.
\vspace{\baselineskip}\newline 
The above injections of one or two positive spectra yielded stable equilibria. 
However, if we inject the last two instabilities at $n = 14$ or $n = 15$, we see the development of chaos in Figure \ref{fig:kt14} and \ref{fig:kt15}.
It is of interest that these modes cause the transition to chaos where the stronger modes $n = 11$ and $n = 12$ did not.
Chaos also ensues if we can also inject the first four wavenumbers and find an unstable periodic solution which disintegrates as time progresses, as shown in Figure \ref{fig:ktleft4}.
This is an interesting comparison with the injection of the most unstable mode, where we see an equilibrium that appears to be potentially unstable initially but then settles.
The contrast here suggests that it is interactions of instabilities rather than their existence which causes chaos.
\vspace{\baselineskip}\newline 
Having analyzed various possibilities for forcing this system, we turn to the perturbed traveling wave TW1 of Cvitanovic et al. 
Since this initial condition has a much smaller length scale $L = 22$, there are just three positive spectra to inject into the system, but we still recover nontrivial behavior.
When wavenumbers are injected one at a time, we recover an equilibrium, and as an example we show the result of injecting the last unstable wave in Figure \ref{fig:cdemaxeq}.
This is a commonality with the previous case, except that in this case all modes yield stable equilibria.
When we consider combinations of the modes, adding $n = 1$ and $n = 3$ results in a relative equilibrium in Figure \ref{fig:cdepersist} that persists for long times without a transition to chaotic behavior.
Some chaos is recovered when $n = 2$ and $n = 3$ are injected simultaneously, as shown in Figure \ref{fig:cdetransition}, but this eventually leads to a periodic orbit.
We also see a continued transition to chaos when all unstable spectra are present with a decreased growth rate with the scaling factors $\epsilon = 0.2$ and $\epsilon = 0.5$.
In the $\epsilon = 0.2$ case, we see the traveling wave start to propagate but fall apart without completing a period.
Going through this example reinforces that the solutions of the KS equation are very sensitive to the energy injection mechanism, so we need just the right forcing to support a given traveling wave or other equilibrium.
\section{Discussion}

\noindent To analyze the Kuramoto-Sivashinksy equation using the metriplectic framework, we removed all positive spectra caused by the energy injection mechanism.
This separation of the forcing and dissipation has accomplished several objectives.
We found that chaotic solutions of the KS equation both dissipate and undergo nontrivial behavior under metriplectic dynamics. 
The entropy prescribed by the UT algorithm increases further when the forcing of the KS equation is present, allowing the consistency of ordered systems with entropy production.
Mode filtering showed that unstable behavior from the KS system can simplify to equilibria and relative equilibria when less instabilities are present.
Even when all unstable modes are reduced in strength, the eventual transition to chaos remains.
Thus our approach provides a unique perspective both in terms of the metriplectic dynamics and the celebrated dynamical systems implications of the KS equation.
\vspace{\baselineskip}\newline Given that much of our analysis depends on a numerical approach to the KS system, we address some limitations and extensions to our approach.
In contrast to past work on the Navier-Stokes-Fourier system, this simulation does not set up an energy-conserving discretization.\cite{BarhamMorrisonZaidni24}
Further analysis could consider using a symplectic exponential method or discrete gradient method which could tightly conserve the energy of our system.
In addition to conservation properties of the KS system, our simulation exploits the continuous Fourier basis to compute the pseudodifferential operator contained in the entropy. 
As a result, our simulation cannot resolve oscillatory shocks observed as traveling wave solutions.\cite{HooperGrimshaw}
To complete an analysis of solutions to the KS equation, we would need to compute $\mathcal{L}$ in a discontinuous Galerkin or other shock capturing scheme.
\vspace{\baselineskip}\newline 
\noindent We conclude by addressing the thermodynamics of the systems underlying the KS equation.
In the decades since the derivations of the KS equation, several of the settings in which it was derived have been shown to be thermodynamically consistent.
To be specific, we know that the incompressible Navier Stokes system is metriplectic, and gravity is a conservative force, so a thin film down a plane is thermodynamically consistent and has an associated entropy.\cite{MaterassiTassi} 
As an additional case, the Vlasov-Maxwell system for two species plasmas with the Landau collision operator is metriplectic.\cite{KrausHirvijoki}
This would include the electron-ion collisions and Landau damping, key components of the LMRT description of the trapped ion mode. 
Though we have observed interesting behavior with the entropy we have defined, an entropy based on eliminating unstable modes is inherently artificial.
The implications of this study would be supported by a comparison of our entropy metric and the physical entropy of these settings.
An alternative asymptotic analysis of these environments may also lead to a reduced model which retains thermodynamic consistency, creating a truly metriplectic KS equation.

\begin{acknowledgments}
\noindent This work was supported by U.S. Department of Energy Grant No. DE-FG02-04ER-54742. WB was supported by the Laboratory Directed Research and Development program of Los Alamos National Laboratory under project number 20251151PRD1. Los Alamos Laboratory Report LA-UR-25-32124. 
\end{acknowledgments}

\section*{Data Availability Statement}

The data that support the findings of this study are available from the corresponding author upon reasonable request.

\nocite{*}
\bibliographystyle{unsrt}
\bibliography{ksthermo}

@article{Morrison98,
  title = {Hamiltonian description of the ideal fluid},
  author = {Morrison, P. J.},
  journal = {Rev. Mod. Phys.},
  volume = {70},
  issue = {2},
  pages = {467--521},
  numpages = {0},
  year = {1998},
  month = {Apr},
  publisher = {American Physical Society},
  doi = {10.1103/RevModPhys.70.467},
  url = {https://link.aps.org/doi/10.1103/RevModPhys.70.467}
}

@article{KuramotoTsuzuki1976,
    author = {Kuramoto, Y. and Tsuzuki, T.},
    title = {Persistent Propagation of Concentration Waves in Dissipative Media Far from Thermal Equilibrium},
    journal = {Progress of Theoretical Physics},
    volume = {55},
    number = {2},
    pages = {356-369},
    year = {1976},
    month = {02},
    issn = {0033-068X},
    doi = {10.1143/PTP.55.356},
    url = {https://doi.org/10.1143/PTP.55.356}
}

@article{BarhamMorrisonZaidni24,
title = {A thermodynamically consistent discretization of 1D thermal-fluid models using their metriplectic 4-bracket structure},
journal = {Communications in Nonlinear Science and Numerical Simulation},
volume = {145},
pages = {108683},
year = {2025},
issn = {1007-5704},
doi = {https://doi.org/10.1016/j.cnsns.2025.108683},
url = {https://www.sciencedirect.com/science/article/pii/S1007570425000942},
author = {W. Barham and P. J. Morrison and A. Zaidni}
}

@article{Nepomnyashchii1974,
author = {Nepomnyashchii, A. A.},
year = {1974},
title =  "Stability of wavy conditions in a film flowing down an inclined plane",
journal = "Fluid Dynamics",
pages = "354-359",
volume = "9",
issue = "3",
issn = "1573-8507",
url = " https://doi.org/10.1007/BF01025515",
doi = "10.1007/BF01025515",
}

@article{Benney66,
    author = {Benney, D. J.},
    title = {Long Waves on Liquid Films},
    journal = {Journal of Mathematics and Physics},
    year = {1966},
    volume = {45},
    pages = {150-155},
    
}

@article{Homsy1974,
    author = {G. M. Homsy},
    title = {Model Equations for Wavy Viscous Film Flow},
    journal = {Lectures in Applied Mathematics},
    volume = {15},
    year = {1974}
}

@article{LMRT,
title = {Nonlinear Saturation of the Trapped-Ion Mode},
  author = {LaQuey, R. E. and Mahajan, S. M. and Rutherford, P. H. and Tang, W. M.},
  journal = {Physical Review Letters},
  volume = {34},
  issue = {7},
  pages = {391-394},
  numpages = {4},
  year = {1975},
  month = {Feb},
  publisher = {American Physical Society},
  doi = {10.1103/PhysRevLett.34.391},
  url = {https://link.aps.org/doi/10.1103/PhysRevLett.34.391}
}

@article{Sivashinsky1977,
title = {Nonlinear analysis of hydrodynamic instability in laminar flames—I. Derivation of basic equations},
journal = {Acta Astronautica},
volume = {4},
number = {11},
pages = {1177-1206},
year = {1977},
issn = {0094-5765},
doi = {https://doi.org/10.1016/0094-5765(77)90096-0},
url = {https://www.sciencedirect.com/science/article/pii/0094576577900960},
author = {G. I. Sivashinsky}
}

@book{KunduCohen,
author = {Kundu, P. K. and Cohen, I. M. and H. H. Hu},
address = {San Diego},
title = {Fluid {M}echanics},
edition = {2nd ed.},
isbn = {0121782514},
lccn = {2001086884},
publisher = {Academic Press},
year = {2002},
}

@article{NicolaenkoGlobal,
title = {Some global dynamical properties of the {K}uramoto-{S}ivashinsky equations: Nonlinear stability and attractors},
journal = {Physica D: Nonlinear Phenomena},
volume = {16},
number = {2},
pages = {155-183},
year = {1985},
issn = {0167-2789},
doi = {https://doi.org/10.1016/0167-2789(85)90056-9},
url = {https://www.sciencedirect.com/science/article/pii/0167278985900569},
author = {B. Nicolaenko and B. Scheurer and R. Temam}
}

@article{HNZOrder,
title = {Order and complexity in the {K}uramoto-{S}ivashinsky model of weakly turbulent interfaces},
journal = {Physica D: Nonlinear Phenomena},
volume = {23},
number = {1},
pages = {265-292},
year = {1986},
issn = {0167-2789},
doi = {https://doi.org/10.1016/0167-2789(86)90136-3},
url = {https://www.sciencedirect.com/science/article/pii/0167278986901363},
author = {J. M. Hyman and B. Nicolaenko and S. Zaleski}
}

@article{GreeneKim,
title = {The steady states of the {K}uramoto-{S}ivashinsky equation},
journal = {Physica D: Nonlinear Phenomena},
volume = {33},
number = {1},
pages = {99-120},
year = {1988},
issn = {0167-2789},
doi = {https://doi.org/10.1016/S0167-2789(98)90013-6},
url = {https://www.sciencedirect.com/science/article/pii/S0167278998900136},
author = {J. M. Greene and J.-S. Kim}
}

@article{CDE10,
author = {Cvitanovi\'{c}, P. and Davidchack, R. L. and Siminos, E.},
title = {On the State Space Geometry of the {K}uramoto–{S}ivashinsky Flow in a Periodic Domain},
journal = {SIAM Journal on Applied Dynamical Systems},
volume = {9},
number = {1},
pages = {1-33},
year = {2010},
doi = {10.1137/070705623},
URL = {https://doi.org/10.1137/070705623}
}

@article{MorrisonUpdike,
  title = {Inclusive curvaturelike framework for describing dissipation: Metriplectic 4-bracket dynamics},
  author = {Morrison, P. J. and Updike, M. H.},
  journal = {Physical Review E},
  volume = {109},
  issue = {4},
  pages = {045202},
  numpages = {22},
  year = {2024},
  month = {Apr},
  publisher = {American Physical Society},
  doi = {10.1103/PhysRevE.109.045202},
  url = {https://link.aps.org/doi/10.1103/PhysRevE.109.045202}
}

@article{ZaidniMorrison,
  title = {Metriplectic four-bracket algorithm for constructing thermodynamically consistent dynamical systems},
  author = {Zaidni, A. and Morrison, P. J.},
  journal = {Physical Review E},
  volume = {112},
  issue = {2},
  pages = {025101},
  numpages = {14},
  year = {2025},
  month = {Aug},
  publisher = {American Physical Society},
  doi = {10.1103/r2lb-xkq6},
  url = {https://link.aps.org/doi/10.1103/r2lb-xkq6}
}

@book{Canuto2006,
author = {Canuto, C. and M. Y. Hussaini and A. Quarteroni and T. A. Zang},
address = {Berlin, Heidelberg},
title = {Spectral {M}ethods : {F}undamentals in {S}ingle {D}omains},
edition = {1st ed.},
isbn = {1-281-34113-4},
publisher = {Springer},
series = {Scientific Computation},
year = {2006},
}

@article{CoxMatthews,
title = {Exponential Time Differencing for Stiff Systems},
journal = {Journal of Computational Physics},
volume = {176},
number = {2},
pages = {430-455},
year = {2002},
issn = {0021-9991},
doi = {https://doi.org/10.1006/jcph.2002.6995},
url = {https://www.sciencedirect.com/science/article/pii/S0021999102969950},
author = {S. M. Cox and P. C. Matthews}
}

@article{KassamTrefethen,
author = {Kassam, A.-K. and Trefethen, L. N.},
title = {Fourth-Order Time-Stepping for Stiff {PDE}s},
journal = {SIAM Journal on Scientific Computing},
volume = {26},
number = {4},
pages = {1214-1233},
year = {2005},
doi = {10.1137/S1064827502410633},
URL = {https://doi.org/10.1137/S1064827502410633}
}

@article{MichelsonStates,
title = {Steady solutions of the {K}uramoto-{S}ivashinsky equation},
journal = {Physica D: Nonlinear Phenomena},
volume = {19},
number = {1},
pages = {89-111},
year = {1986},
issn = {0167-2789},
doi = {https://doi.org/10.1016/0167-2789(86)90055-2},
url = {https://www.sciencedirect.com/science/article/pii/0167278986900552},
author = {D. Michelson}
}

@article{HooperGrimshaw,
title = {Travelling wave solutions of the {K}uramoto-{S}ivashinsky equation},
journal = {Wave Motion},
volume = {10},
number = {5},
pages = {405-420},
year = {1988},
issn = {0165-2125},
doi = {https://doi.org/10.1016/0165-2125(88)90045-5},
url = {https://www.sciencedirect.com/science/article/pii/0165212588900455},
author = {A. P. Hooper and R. Grimshaw}
}

@article{FrischCells,
title={Viscoelastic behaviour of cellular solutions to the {K}uramoto-{S}ivashinsky model}, volume={168},
DOI={10.1017/S0022112086000356},
journal={Journal of Fluid Mechanics},
author={Frisch, U. and She, Z. S. and Thual, O.},
year={1986},
pages={221–240}
}

@article{KrausHirvijoki,
    author = {Kraus, M. and Hirvijoki, E.},
    title = {Metriplectic integrators for the {L}andau collision operator},
    journal = {Physics of Plasmas},
    volume = {24},
    number = {10},
    pages = {102311},
    year = {2017},
    month = {10},
    issn = {1070-664X},
    doi = {10.1063/1.4998610},
    url = {https://doi.org/10.1063/1.4998610}
}

@article{MaterassiTassi,
title = {Metriplectic framework for dissipative magneto-hydrodynamics},
journal = {Physica D: Nonlinear Phenomena},
volume = {241},
number = {6},
pages = {729-734},
year = {2012},
issn = {0167-2789},
doi = {https://doi.org/10.1016/j.physd.2011.12.013},
url = {https://www.sciencedirect.com/science/article/pii/S0167278911003708},
author = {M. Materassi and E. Tassi}
}

@article{2002VDSMaschke,
title = {Hamiltonian formulation of distributed-parameter systems with boundary energy flow},
journal = {Journal of Geometry and Physics},
volume = {42},
number = {1},
pages = {166-194},
year = {2002},
issn = {0393-0440},
doi = {https://doi.org/10.1016/S0393-0440(01)00083-3},
url = {https://www.sciencedirect.com/science/article/pii/S0393044001000833},
author = {A. J. {van der Schaft} and B. M. Maschke}
}

@article{Eberard2007,
title = {An extension of {H}amiltonian systems to the thermodynamic phase space: Towards a geometry of nonreversible processes},
journal = {Reports on Mathematical Physics},
volume = {60},
number = {2},
pages = {175-198},
year = {2007},
issn = {0034-4877},
doi = {https://doi.org/10.1016/S0034-4877(07)00024-9},
url = {https://www.sciencedirect.com/science/article/pii/S0034487707000249},
author = {D. Eberard and B. M. Maschke and A. J. {van der Schaft}}
}

@article{Morrison1986,
title = {A paradigm for joined {H}amiltonian and dissipative systems},
journal = {Physica D: Nonlinear Phenomena},
volume = {18},
number = {1},
pages = {410-419},
year = {1986},
issn = {0167-2789},
doi = {https://doi.org/10.1016/0167-2789(86)90209-5},
url = {https://www.sciencedirect.com/science/article/pii/0167278986902095},
author = {P. J. Morrison}
}

@article{MetRelaxEq,
title = {Metriplectic Relaxation to Equilibria},
journal = {arXiv:2506.09787v1 [math-ph]},
volume = {},
number = {},
pages = {},
year = {2025},
issn = {},
doi = {https://doi.org/10.48550/arXiv.2506.09787},
url = {https://arxiv.org/abs/2506.09787},
author = {C. Bressan and M. Kraus and O. Maj and P. J. Morrison}
}

@article{Morrison1984,
    author = {P. J. Morrison},
    title = {Some Observations Regarding Brackets and Dissipation},
    journal = {Center for Pure and Applied Mathematics Report PAM-228, University of California, Berkeley},
    year = {1984},
    note = {Available at arXiv:2403.14698v1 [math-ph], 15 Mar 2024}
}

\end{document}